\pdfoutput=1
\documentclass[%
  prb,%
  letterpaper,%
  twocolumn,%
  showpacs,%
  floatfix,%
  superscriptaddress,%
  longbibliography,
]{revtex4}

\RequirePackage{graphicx,amsmath,amssymb,bm}
\RequirePackage{hyperref}

\newcommand{\Fref}[1]{Fig.~\ref{#1}}

\renewcommand{\eqref}[1]{eq.~(\ref{#1})}

\usepackage{booktabs} 

\usepackage[usenames,dvipsnames]{xcolor}
\usepackage{soul}

\begin{document}

\title[Valley notch filter in a graphene strain superlattice]{ Valley notch 
filter in a graphene strain superlattice: Green's function and machine 
learning approach}

\author{V. Torres}
\affiliation{Instituto de F\'isica, Universidade Federal Fluminense, 
Niter\'oi, Av. Litor\^{a}nea sn 24210-340, RJ-Brazil}
\affiliation{MackGraphe\,--\,Graphene and Nano-Materials Research Center,
Mackenzie Presbyterian University, Rua da Consola\c{c}\~{a}o 896, 01302-907,
S\~{a}o Paulo, SP, Brazil}
\author{P. Silva}
\affiliation{MackGraphe\,--\,Graphene and Nano-Materials Research Center,
Mackenzie Presbyterian University, Rua da Consola\c{c}\~{a}o 896, 01302-907,
S\~{a}o Paulo, SP, Brazil}
\author{E. A. T. de Souza}
\affiliation{MackGraphe\,--\,Graphene and Nano-Materials Research Center,
Mackenzie Presbyterian University, Rua da Consola\c{c}\~{a}o 896, 01302-907,
S\~{a}o Paulo, SP, Brazil}
\author{L. A. Silva}
\affiliation{School of Computing and Informatics \& Graduate Program 
in Electrical Engineering and Computing, Mackenzie Presbyterian University}
\author{D. A. Bahamon}
\email{dario.bahamon@mackenzie.br}
\affiliation{MackGraphe\,--\,Graphene and Nano-Materials Research Center,
Mackenzie Presbyterian University, Rua da Consola\c{c}\~{a}o 896, 01302-907,
S\~{a}o Paulo, SP, Brazil}

\begin{abstract}
The valley transport properties of a superlattice of out-of-plane 
Gaussians deformations are calculated using a Green's function and a Machine 
Learning approach. Our results show that periodicity significantly improves the 
valley filter capabilities of a single Gaussian deformation, these manifest 
themselves in the  conductance as a sequence by  valley filter plateaus. We 
establish that the physical effect behind the observed valley notch filter is  
the coupling between counter-propagating transverse modes; the complex 
relationship between the design parameters of the superlattice and the valley 
filter effect make difficult to estimate in advance the valley filter 
potentialities of a given superlattice. With this in mind, we show that a Deep 
Neural Network can be trained to predict   valley polarization with a precision 
similar to the  Green's function but with much less computational effort.
\end{abstract}


\maketitle

\section{Introduction}

The generation, control and detection of electrons from different  valleys 
is called valleytronics, the valley quantum  number  naturally appears in 
periodic solids with  degenerated local minima and maxima at inequivalent  
points of the Brillouin zone \cite{Valley2D}. The idea of manipulating the 
valley  to store and to process information is not new 
\cite{PhysRevLett.40.472,Thompson:2004kq,PhysRevB.74.155436}, however, there is 
renewed interest in this field due to the appearance of 2D materials. One atom 
thick layers with hexagonal lattices such as graphene and transition metal 
dichalcogenides offer two valleys (K and K') well separated in momentum space 
that can be accessed by optical \cite{PhysRevB.77.235406,Cao:2012fk}, magnetic 
\cite{PhysRevLett.113.266804, PhysRevLett.114.037401} and mechanical means 
\cite{nl400872q,7b01663,carrillo2016strained,da_Costa_2017}. Although many of 
these approaches have made great success producing valley currents, they require 
high-quality samples with perfect alignment between the layer and the substrate  
\cite{Gorbachev448,Lee:2016kx}. On the other hand, inhomogeneous mechanical 
deformations such as bubbles and ripples routinely appear in 
graphene\cite{Levy544};   these are seen by electrons in opposite valleys as 
regions with opposite polarity pseudo-magnetic fields. This attribute has been 
used in devices with one graphene 
bubble\cite{PhysRevLett.117.276801,Carrillo-Bastos:2018kq,Stegmann,
Milovanovic:2018fj,myoung2019,milovanovic2016strain,Milovanovi__2016} to show 
separation of valley currents and valley filtering; unfortunately, the observed 
effects require fine-tuning of the energy, defined height/width ratio of the 
bubble, high values of strain, narrow contacts, location of the nanobubble near 
to the right contact and crystalline orientation.  Clearly, the proposals with a 
single graphene bubble present serious disadvantages to efficiently generate and 
detect valley currents\cite{PhysRevB.98.165437}.

With the objective of improving the valley filtering capabilities observed in 
a single graphene nanobubble, in this study we focus on the electronic and 
valley transport properties of a 1D superlattice of graphene Gaussian  
out-of-plane deformations in a zigzag graphene nanoribbon. It is well known 
that one-dimensional periodic potentials modify the electronic properties of 
bulk graphene producing anisotropic charge transport\cite{Park:2008yq}, 
additional Dirac points\cite{PhysRevLett.103.046809} and a tunable band 
gap\cite{Wang:2010sf,Andrade:2019zl}. In addition, in graphene nanoribbons 
periodicity couples transverse modes promoting selective  reflection 
\cite{PhysRevB.79.155409}. From the experimental point of view, it has been 
shown the impressive capacity of depositing graphene on nanopatterned 
substrates\cite{Jiang:2017hl,Gill:2015fp,Hinnefeld:2018fv,banerjee2019strain}; 
local measurements of the electronic properties have shown the appearance of 
pseudo-Landau levels in the strained regions  providing direct evidence of the 
formation of strain superlattices\cite{Jiang:2017hl}.

Our study shows that periodicity really enhances the valley filter 
capabilities of the Gaussian out-of-plane deformation. Using the lattice 
Green's function and the wave function matching technique we identify that the 
combined effects of strain and periodicity give rise to a notch valley filter 
effect, the selective rejection of electrons in one valley is originated by the 
diagonal and non-diagonal coupling of counter-propagating modes. The main 
significant advantage of the Gaussian superlattice is that the observed valley 
filter effect emerges in wider energy regions with low height/width ratios. In 
general terms,  it is difficult to predict with any certainty the number, 
bandwidth and energy location of the valley filters. To estimate them we use 
Machine Learning\cite{Zdeborova:2017rc},  this alternative approach has recently 
emerged as a new tool to design the properties of different physical 
systems\cite{PhysRevLett.115.205901,PhysRevLett.121.255304,Cubuk:2017nx,
PhysRevB.97.115453}. Concretely, we show that a Deep Neural Network predicts the 
valley transport properties with nearly the same accuracy of the Green's 
functions, with this new tool we explore the configuration space to extract the 
superlattice with the smallest number of Gaussians and strain that maximizes 
valley transport.

\section{Modelling valley transport using Green's functions}
\begin{figure}
\centering
\scalebox{0.15}{
\includegraphics{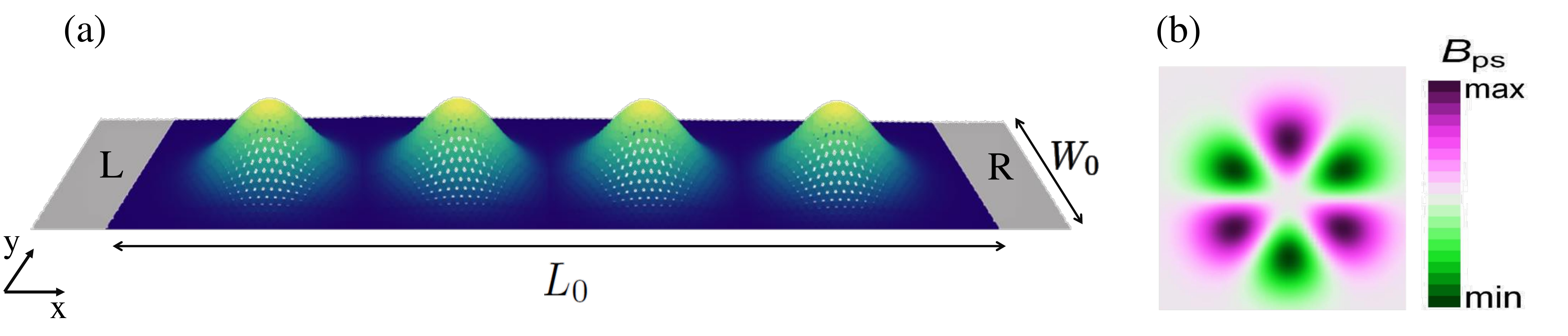}}
\caption{ (a) Schematic representation of the two-terminal system. The 
central part of the ribbon has dimensions $ L_0 \times W_0$ and the zigzag 
direction is along to the $x$ axis.  The terminals are labeled $L$ and $R$. The 
system is deformed by an array of $N_G$ out-of-plane Gaussian deformations in 
series. (b) Profile of the pseudo-magnetic field for one Gaussian bump in  $K$ 
valley.}
\label{figure1}
\end{figure}

We consider a graphene section with zigzag edges and dimensions $L_0 \times 
W_0$ connected to two pristine semi-infinite zigzag graphene nanoribbons (see 
\Fref{figure1}(a)); the electronic and transport properties  of this system are 
calculated within the nearest neighbor tight-binding Hamiltonian. The effect of 
the mechanical deformation is included through the modification of the hopping 
energy between sites $i$ and $j$:
 
\begin{equation}
t_{ij}=t_{0}\exp\left[-\beta\left(\frac{l_{ij}}{a_c}-1\right)\right], \label{
hopping01}
\end{equation}

\noindent where $t_0=2.7$ eV is the nearest neighbour hopping of graphene 
without any deformation, $a_{c} = 1.42$ $\AA \ $ is the carbon-carbon distance,  
$\beta = (\partial \ln t/\partial \ln a)_{a=a_{c}}\approx 3$ and  $l_{ij}$ is 
the new interatomic distance under strain. In the central region we include an 
array of  \textit{N} out-of-plane Gaussian deformations given by:

\begin{equation}
h(x,y) = \sum_n^N A_n \exp
\left[-\frac{(x - x_n)^2+(y - y_n)^2}{b_n^2}\right],
\end{equation}

\noindent where $(x_n,y_n)$ is the center of the \textit{n-th} Gaussian bump, 
$A_n$ is the height and $b_n$ is the width. For simplicity we take the same 
height and  width for all \textit{N} Gaussian deformations ($A_n = A$ and $b_n = 
b$). 
Electrons in graphene see mechanical deformations as regions with a  gauge 
field \cite{vozmediano2010gauge} defined by:

\begin{equation}
\boldsymbol{A}_{ps}=\frac{\beta\hbar 
v_F}{2a_{c}}\left(\epsilon_{xx}-\epsilon_{yy} ,
-2\epsilon_{xy} \right).
\end{equation}

\noindent where $v_F$ is the Fermi velocity and $\epsilon_{ij}$ is the 
strain tensor. The typical pseudo-magnetic field $B_{ps}=\nabla \times A_{ps}$ 
for $K$ valley generated by one Gaussian deformation is shown in the Figure 
\ref{figure1}(b), for the opposite valley the polarity is reversed.


In the Landauer--B\"uttiker formalism, the two terminal conductance depends on 
the transmission probability that one electron injected along the left edge of 
the scattering region or device will transmit to the right edge. Here, we 
consider as scattering region the central section where the out-of-plane 
Gaussian deformations  are included (see \Fref{figure1}(a)).
It is noteworthy that to access the valley degree of freedom a 
mixed representation is required. We have to consider the transverse modes of 
the contacts as well as the modification of the hopping energies between 
neighboring sites in the lattice.  With this in mind,  we calculate the 
transmission probability by matching the wave functions in the scattering region 
to the Bloch modes of the pristine contacts 
\cite{ando1991quantum,khomyakov2005conductance}. The transmission matrix 
elements for the incoming  mode $m$ in the left contact and outgoing mode $n$ 
in the right contact can be expressed as:

\begin{equation}
\begin{split}
&
\tau_{n,m}=\sqrt{\frac{v_{R,n}}{v_{L,m}}}\left[{u}^{\dagger}_{R,n}G_{RL} Q_{0}  
u_{L,m}\right]
\end{split}
\label{tnm}
\end{equation}

\noindent where ${u}_{R,n}$ (${u}_{L,m}$)  is the outgoing $n$ (incoming 
$m$) mode, $v_{R,n}(v_{L,m})$ is the velocity of the mode ${u}_{R,n}$ 
(${u}_{L,m}$), $G_{RL}$ is the  Green's function and $Q_{0}$ is the source term. 
Using the mode matching method we can easily split the conductance, $G=(2e^2/h) 
\sum_{m,n}\mid \tau_{n,m}\mid^2$,   into their valley components  $G =(2e^2/h) 
\left [ \gamma_{K} + \gamma_{K'} \right ]$, where the valley transmission 
$\gamma_{K(K')}=\sum_{m \in K(K'),n}\mid \tau_{n,m}\mid^2$ is related to the 
transmission probability that the incoming electron in mode $m$ in $K$ or $K'$ 
valley  is scattered to mode $n$ in $K(K')$ valley . Once the transmission per 
valley is calculated the polarization is easily determined by:

\begin{equation}
\begin{split}
P_{K (K^{'})} =\frac{ \gamma_{K (K')}}{ (\gamma_{K} + \gamma_{K^{'}})} 
\end{split}
\end{equation}

\subsection{Single Gaussian deformation} 

\begin{figure}
\centering
\scalebox{0.44}{\includegraphics{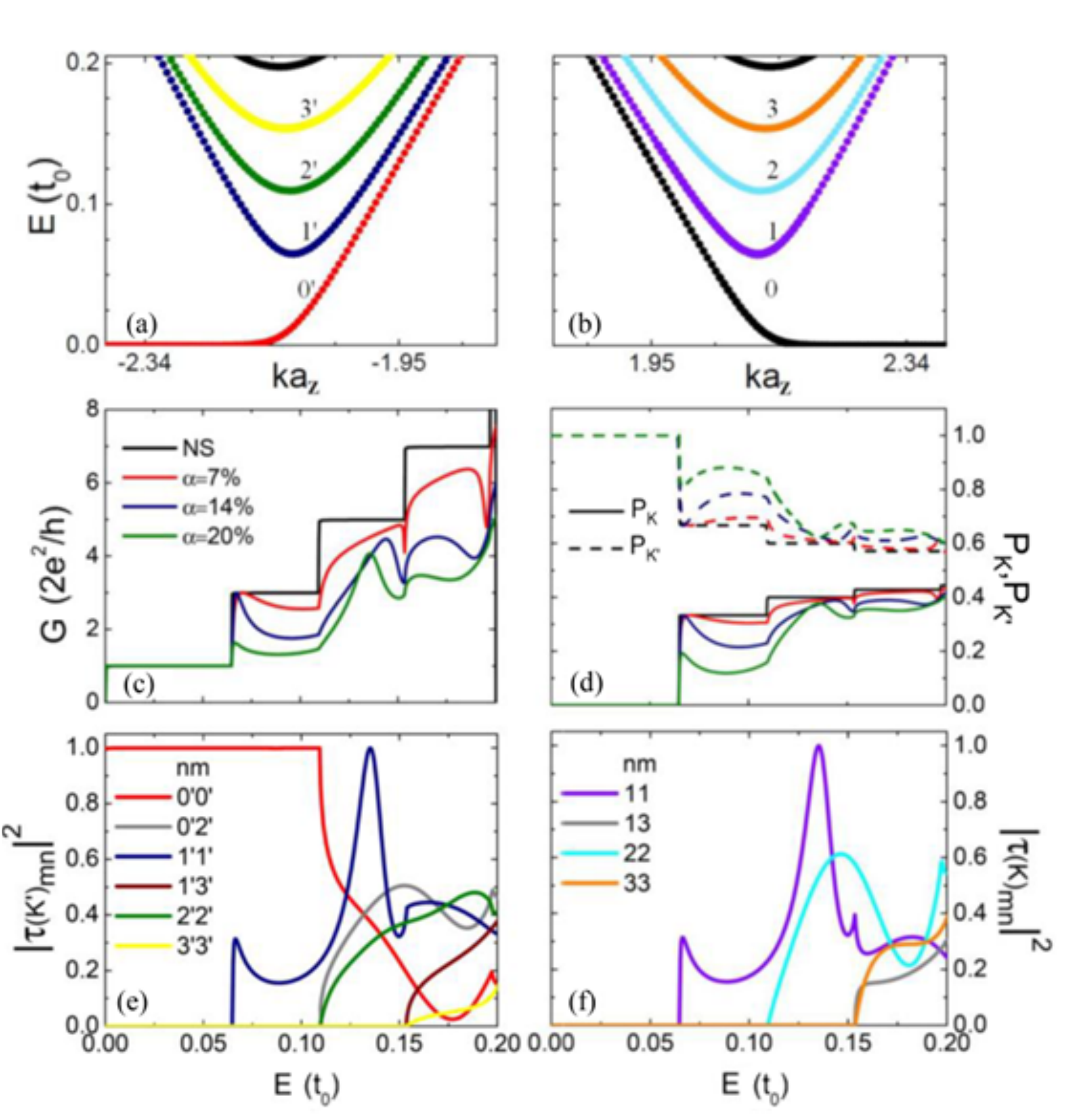}}
\caption{Band structure for a zigzag graphene nanoribbon of width $ W_0 \approx 
14.8 $ nm around  (a) $K'$ valley  and (b) $K$ valley. (c) Conductance and (d) 
Valley polarization $P_{K'}$ and $P_K$ in the presence of one out-of-plane 
Gaussian deformation for different values of the strain parameter $\alpha$.   
Transmission probability for the low energy modes in (d) $K'$ valley  and (e) 
$K$ valley  produced by one Gaussian bump with $\alpha=20\%$.
} 
\label{figure2} 
\end{figure}

The transport properties  of a single graphene nanobubble  have been 
previously investigated\cite{PhysRevLett.117.276801,Carrillo-Bastos:2018kq,
Stegmann,Milovanovic:2018fj,carrillo2014gaussian}, in this section for 
completeness reasons we start presenting the main results for conductance and  
valley polarization, then, we use the wave function matching technique to gain 
physical insight into the origin of the observed valley effect.  We consider a 
scattering region with dimensions $L_0 \approx 12.8$ nm and $ W_0 \approx 14.8 $ 
nm with a Gaussian bump of fixed-parameter  $b = 22a_c = 3.12~\text{nm}$ 
localized exactly at the center of the system. The conductance 
calculation requires the transmission probability $\tau_{n,m}$ between 
transverse modes, we therefore identify in  \Fref{figure2}(a)-(b) the low energy 
transverse mode dispersion around $K$ and $K'$ valley  in the contacts (we use 
the primitive unit cell with $a_z = \sqrt{3}a_c$). \Fref{figure2}(c) shows the 
evolution of the conductance for different values of the parameter $ \alpha = (A 
/ b)^2 $, which determines the strain generated by the bubble. For bubbles with 
low curvature, the conductance presents sharp dips at the outset of a new 
conductance plateau, larger curvatures (increase in the value of $A$) widen the 
dips and degrade the overall conductance.  From the point of view of valley 
transport, it has been observed (see \Fref{figure2}(d)) that  a single Gaussian 
deformation induces some degree of valley polarization for the lower energy 
transverse modes, obviously  the fully valley polarized  zigzag edge state 
plateau ($P_{K'}=1$) is disregarded. 


It is worth stressing  that the electric current in the contacts is carried  by 
independent transverse modes, so the observed valley imbalance introduced by the 
Gaussian deformation is, in the end, a mode mixing effect. We can have a clear 
picture of the intervalley and intravalley scattering processes using 
\eqref{tnm}, this is done for the deformation with $\alpha = 20\%$ in 
\Fref{figure2}(e)-(f) where $|\tau_{n,m}|^2$ is plotted for the low energy modes 
in $K'$ and $K$ valley respectively. The transmission probabilities reveal that 
the valley filter effect is entirely created by the fully valley polarized 
zigzag edge state mode, $|\tau_{0',0'}|^2 = 1$ until the onset of the second 
transverse mode at $E_{2'(2)} = 0.11t_0$. In general, there is no such a thing 
as electrons in one valley are transmitted through the deformation while 
electrons in the other valley are reflected\cite{PhysRevB.82.205430}, transverse 
modes in $K'$ valley   are scattered in a similar way  as transverse modes in 
the opposite $K$ valley. The valley filter effect is  purely and simply observed 
because $K'$ valley  has an extra mode.

It is also interesting to look at the peak $|\tau_{1',1'(1,1)}|^2 = 1$ at 
$E=0.135t_0$ and the dip in $|\tau_{0',0'}|^2 = 0$ at $E=0.177t_0$, both 
energies signal the presence of quasi-bound states\cite{carrillo2014gaussian} 
below the  third  ($E_{3'(3)} = 0.154t_0$)  and fourth ($E_{4'(4)} = 0.197t_0$) 
transverse modes. We find that the energy of these new states  ($E_4-0.177t_0 = 
0.02t_0$ and $E_3-0.135t_0 = 0.02t_0$) decreases as $\alpha^{2.43}$. Although 
the existence of the quasi-bound state is not noticed in the conductance plot, 
different lineshapes for  transmission probabilities of independent modes and 
the suppression of intravalley scattering for others ($|\tau_{0',1'}|^2 
=|\tau_{1',2'(1,2)}|^2 = |\tau_{2',3'(2,3)}|^2 =0$) show that the Gaussian 
deformation scatters differently electrons on transverse modes in the same 
valley.  The observed features are not  absolutely governed by the 
pseudo-magnetic field ($B_{ps} \approx 515~\text{T}$ for $\alpha = 20\%$) or the 
geometry of the scatterer, note that there are signatures on the conductance or 
transmission for energies below  $E_{ps} =  \hbar v_F/\ell_B \approx 0.19t_0$ 
and $E_{\text{scatt}} = \hbar v_F\pi/b \approx 0.2t_0$.

In short, we have shown that for one out-of-plane Gaussian deformation the 
valley imbalance is entirely created by the zigzag edge states mode, 
unfortunately, this fact seriously restricts its use as  valley filter device.   
On the other hand, we also highlight the multimode nature of the electronic 
transport, although at first glance this may seem undesirable, in periodic 
structures  mode mixing results in the formation of mini-stopbands  
\cite{PhysRevLett.71.137,PhysRevB.79.155409}. These local  gaps in the band 
structure reject specific transverse modes; in the next section  using  a 
sequence of out-of-plane Gaussian deformations, we explore this notch filter 
effect in  the valley transport domain. 

\subsection{Strain Superlattice: 1D Gaussian Chain.} 
\label{sec:1DGC}

\begin{figure}
\centering
\scalebox{0.4}{\includegraphics{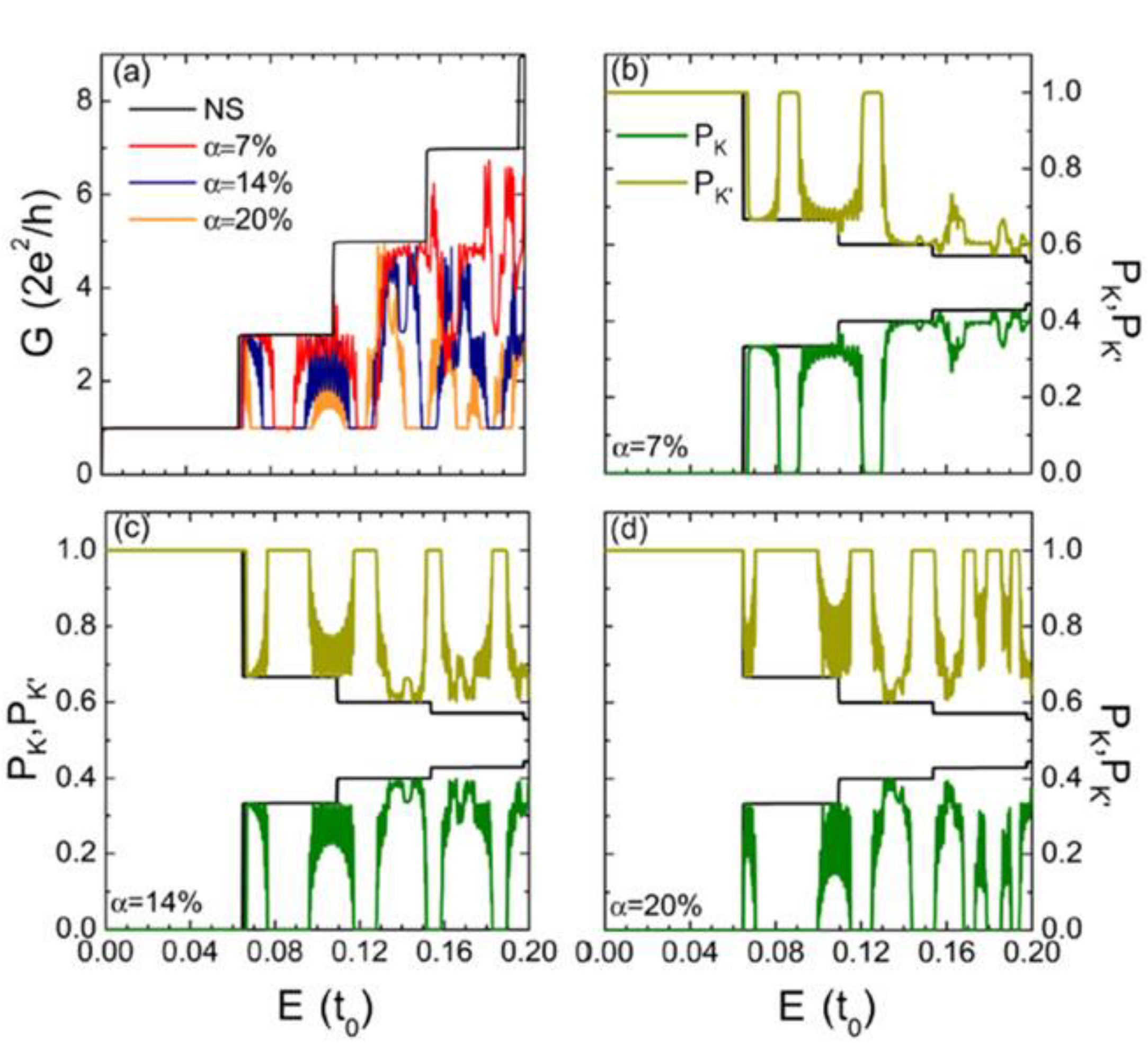}}
\caption{ (a) Conductance for a zigzag graphene nanoribbon with an arrangement 
of $N_G =15$ Gaussian deformations and $ W_0 \approx 14.8 $ nm. Valley 
polarization $P_{K'}$ and $P_K$ for the same system with (b) $\alpha = 7\%$, (c) 
$\alpha = 14\%$ and (d) $\alpha = 20\%$
} 
\label{figure3} 
\end{figure}

We arrange $N_G$ Gaussian bumps along the $x$ direction, the centers of the 
bumps are separated  by $d = 52\sqrt{3}a_c \approx 12.79~\text{nm}$, the width 
of the bumps $b = 22 a_c = 3.12~\text{nm}$ and the width of the zigzag 
nanoribbon $W_0 \approx 14.8 $ nm are the same as the previous section. The 
center of the first(last) Gaussian is located at $d/2$ from the left(right) 
contact, so the length of the central region is $ L_0 =N_Gd$.  To promote strong 
mode coupling and subsequently the formation of mini-stopbands from the very 
beginning we start with a large number of Gaussians  ($N_G = 15$); in 
\Fref{figure3}(a) we follow the evolution of  the conductance of this 1D 
Gaussian chain for different values of $\alpha$. The conductance shows strong  
modulation indicating  the formation of well-defined minibands,  however, 
looking at the transport signatures it is clear that there are two distinct 
features. The first kind produces rapidly oscillating conductance peaks, 
illustrating mode mixing, the number of peaks in one band is equal to the number 
of Gaussian deformations $N_G$.  We do not show but LDOS plots of the peaks in 
these bands present high electronic density precisely on the Gaussian 
deformation region, therefore, these bands are originated by the coupling of the 
quasi-bound states localized on individual bubbles. Highly localized states are 
weakly coupled and result in narrow bands for $\alpha=20\%$, while less 
localized wave functions are strongly coupled and produce wide bands for 
$\alpha=7\%$. On the other hand, the second element in the conductance produces 
perfectly quantized conductance plateaus  ($G=G_0=2e^2/h$), as the value of 
$\alpha$ increases, so do the number and the width of them. These plateaus are 
fully valley polarized in $K'$ valley  as shown in \Fref{figure3}(b)-(d); 
notably, the combination of strain, periodicity and mode mixing produces a 
highly efficient valley filter. 


Although both types of Conductance signatures (resonant peaks and plateaus) 
show integer values of $G_0$, the transmission probabilities of individuals 
modes are not totally quantized, the quantization  is the result of intra-valley 
mode mixing. Just like in the single Gaussian deformation exposition 
$|\tau_{0',0'}|^2 = 1$ for $E \leq E_{2'(2)} = 0.11t_0$, in this way, the first 
two plateaus $P_{K'} = 1$ seen in \Fref{figure3}(b)-(d) are produced by the 
zigzag edge states traverse mode. While higher energy valley polarized plateaus 
are created by incoming electrons in mode $m=0'~\text{and}~2'$  scattered into 
mode $n=2'$ of the  $K'$ valley, in all of the cases studied we found  
$|\tau_{2',2'}|^2 > 0.4$. This is another strength of the 1D Gaussian 
superlattice valley filter  because higher energy transverse modes are less 
affected by edge roughness. 


\begin{figure}
\centering
\scalebox{0.4}{\includegraphics{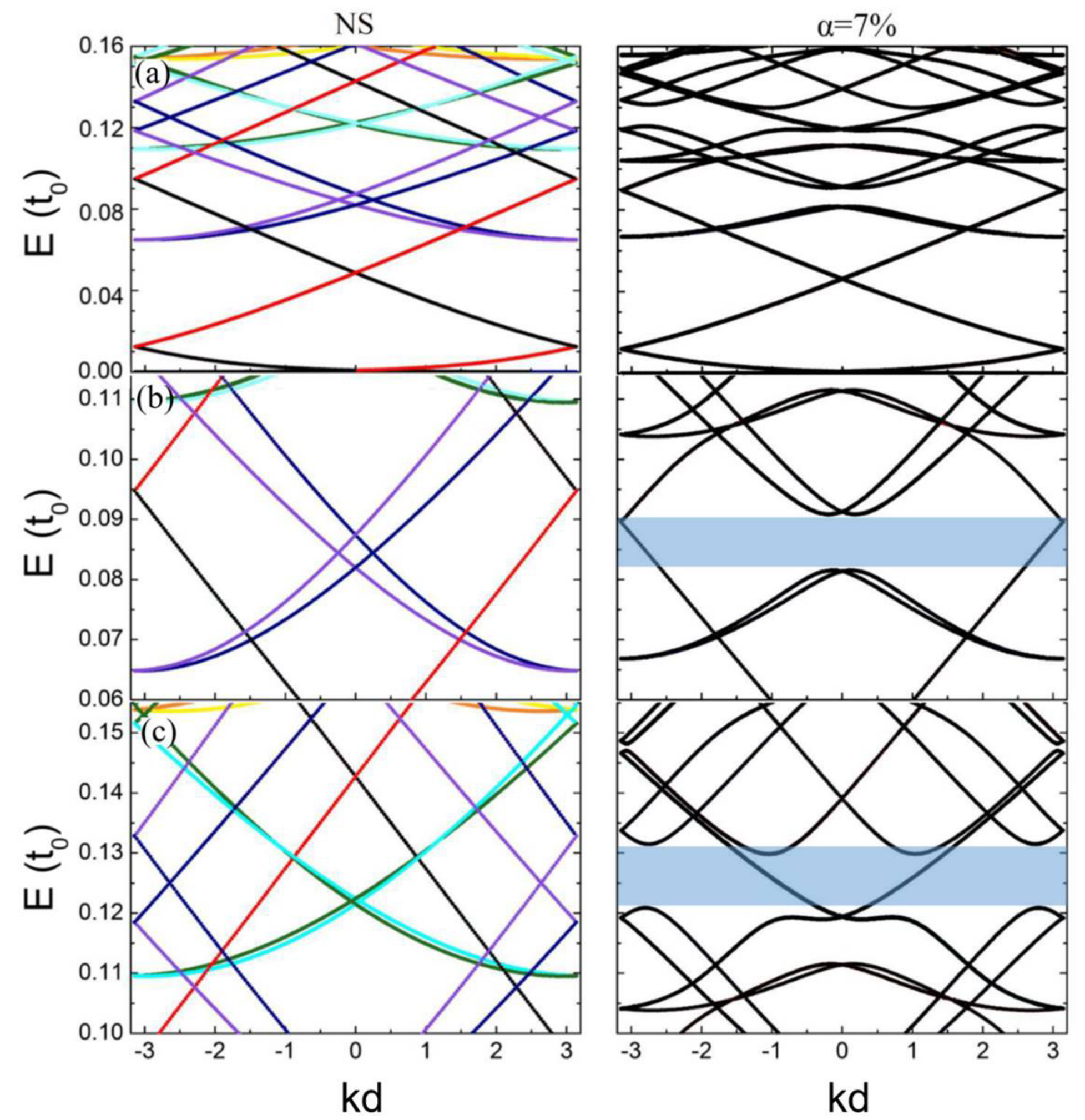}
}
\caption{ Left panel: folded band structure of a zigzag graphene nanoribbon 
without deformation and supercell lattice constant $d =  52a_z$ and $W_0 \approx 
14.8 $ nm, the colours identify the transverse modes of the primitive unit cell 
presented in \Fref{figure2}(a)-(b) . Right panel: band structure of an infinite 
1D chain of out-of-plane Gaussian deformations with $\alpha = 7\%$. Close-up at 
the energy region corresponding to  the second (b) and  third valley (c) filter 
plateau $P_{K'} = 1$ and $P_K =0$ in \Fref{figure3}(b).
}
\label{figure4} 
\end{figure}


In order to further characterize the valley filter effect, we calculated the 
band structure of the corresponding infinite 1D chain of Gaussian bumps with 
$\alpha = 7\%$. The result is presented in the right panel of \Fref{figure4}, a 
close-up at the energy region of the second and third valley filter plateau is 
shown in  panels (b) and (c), the shaded regions mark the exact location and 
width of the plateaus. We also plot as a reference in the left panels the folded 
band structure of the same nanoribbon without out-of-plane deformations, the 
colors correspond to the independent transverse modes of the primitive unit cell 
of \Fref{figure2}(a)-(b). Due to the folding, the unstrained supercell shows 
crossings between modes in the same  or in the opposite valley, it is important 
to note that the periodic potential could lift the degeneracy at these points 
\cite{PhysRevB.79.155409}. A comparison between the left and right panels of 
\Fref{figure4} suggests  that  mini-stopbands (anticrossings) appear as a result 
of the coupling between  counter-propagating modes in the same valley. In 
general terms, contra-directional modes $m$ and $n$ get coupled when  $k_m-k_n 
=\Delta k_{m,n} = \pm \ell G$, where $\ell$ is an integer, $G=2\pi/d$ , $k_m = 
\sqrt{(E/\hbar v_F)^2 - q_m^2}$ and $q_m = \pi(m+\frac{1}{2})/W_0$. Diagonal 
couplings ($m=n$)  generally appear at the Brillouin zone boundary 
($k_m=\pi/d$); however, we are observing diagonal (anti)crossings at different 
zone values. To confirm the formation of standing waves, we calculated the 
energy ($E_{mm}^{\ell}$) where the Bragg condition is fulfilled for independent 
modes:

\begin{equation}
 E_{mm}^{\ell} = \frac{3t_0a_c}{2}\sqrt{\left( \frac{\ell \pi}{d}\right)^2 
+ \left( \frac{\pi (m+\frac{1}{2}) }{W_0} \right)^2}
\end{equation}
\label{Eq:BraggE}

It is clearly seen in the left panel of \Fref{figure4}(b)-(c) crossings at 
$E_{1'1'(11)}^{1} = 0.085t_0$ and $ E_{1'1'(11)}^{2} = 0.124t_0$ for the 
blue(purple) band and anticrossings in the right panel around the same energies. 
Further, the mini-stopband opened by the Bragg reflection of modes $m=1',1$ is 
the physical effect behind the formation of the second valley filter plateau, 
this is concurrent with the mode transmission probability  $|\tau_{0',0'}|^2 = 
1$ calculated in the previous section. The third valley filter plateau is also 
originated by the coupling of counter-propagating modes, however, in this case 
there are diagonal ($m=1'(1), 2'(2)$) and off-diagonal ($m = 0'(0)$ and 
$n=2'(2)$) couplings involved. Based on the previous analysis one may think that 
periodicity alone promotes valley filter. However, this is not the case, the 
increase in the value of $\alpha$ strengthens  the couplings between the low 
energy modes enlarging the anticrossings, and therefore the width of the firsts 
valley filter plateaus. It also fosters the  mixing of higher energy modes 
producing additional plateaus.


To quantify the valley filter capability of the 1D Gaussian superlattice, we 
define

\begin{equation}
\begin{split}
Q_{K^{'}} = \left[ \frac{E(P_{K^{'}} > 0.98)- E_{\text{edge}}}{E_T}\right] 
\times 100 , 
\end{split}
\label{eq:Q}
\end{equation}

\noindent where $E(P_{K^{'}} > 0.98)$ are the energy points with valley 
polarization  $P_{K^{'}} > 0.98$, $E_{\text{edge}}$ is the energy  of the fully 
valley polarized zigzag edge states ($E_{\text{edge}} =q_1\frac{3a_c}{2}t_0 = 
\left[\frac{3\pi}{2W_0}\right]  \frac{3a_c}{2}t_0 $) and $E_T$ is the  energy 
bandwidth (in our case $E_T = 0.2t_0$). For the supperlattices presented in 
\Fref{figure3}  we obtain $Q_{K^{'}} = 9.1\%$ for $\alpha = 7\%$, $Q_{K^{'}} = 
22.5\%$ for $\alpha = 14\%$ and $Q_{K^{'}} = 33.3\%$ for $\alpha = 20\%$. These 
numbers confirm  that the Gaussian out-of-plane superlattice offers significant 
advantages  over the single deformation. First, The valley filter effect does 
not require a fine  tuning of the energy since the filter appears at different 
and wider energy regions. Second, the superlattice demands  low values of strain 
($\alpha$). Third, the superlattice does not depend on perfect zigzag edges 
given that low energy transverse modes get coupled and generate valley filters. 
$Q_{K^{'}}$ allows, not only to quantify the effectiveness of the filter  but 
also the impact of the superlattice parameters on its performance. To that end, 
we modify, one at a time, the values of the height ($A$), the width ($b$) and 
the distance ($d$)  for the superlattice presented in \Fref{figure3}(c) ($A =  
1.17~\text{nm}$, $b=3.12~\text{nm}$, $d = 12.79~\text{nm}$ and  and $N_G = 15$). 
The results are summarised in table \ref{tab:Q}, where $Q_{K^{'}}^u$ is the 
valley filter capability when the value of $u=A,b,d$ is reduced/increased 
($\downarrow/\uparrow)$  by 10, 20 and 30\%. In order to understand the 
behaviour of $Q_{K^{'}}^u$, note that the values of  $A$ and $b$ are related 
with the strength and the  extend of the PMF ($|B_{ps}| \propto \alpha$). A 
contraction of 10, 20 and 30\% in $A$, in fact, is a reduction of  19, 36 and 
51\% in the strength of the PMF. The weakening of the PMF  produces a less 
effective coupling between counter-propagating  modes compressing the value of   
$Q_{K^{'}}^A$.  For $b$, although the region with the PMF is shrunk, the process 
is reversed and  the strength of the  PMF is risen by 23, 56 and 100\%. Thus, 
$Q_{K^{'}}^b$ first rises when the  electrons flowing from the left to the right 
contact sense a  higher PMF, then declines when the PMF region is too narrow 
than only few electrons are filtered. Notwithstanding the reductions  of 
$Q_{K^{'}}$ observed in table \ref{tab:Q},  the 1D Gaussian superlattice 
continues to be a more effective valley filter than the single Gaussian bubble 
for similar $A$ and $b$ parameters 
\cite{milovanovic2016strain,Milovanovi__2016}. Regarding the inter Gaussian 
deformation distance, we observed that in general,   larger  $d$ weakens the  
cascaded configuration of the filter. However, oscillations in $Q_{K^{'}}^d$ can 
be seen given that additional transverse modes can get coupled.

\begin{table}[b]
\centering
\begin{tabular}{c|c|c|c}
\toprule
\textbf{Variation (\%)} &     \textbf{$Q_{K^{'}}^A$~($\downarrow$)} &    
 \textbf{$Q_{K^{'}}^b$~($\downarrow$)} &   \textbf{$Q_{K^{'}}^d$~($\uparrow$)}  
\\
\midrule

0        &  22.5 &  22.5 &  22.5  \\
10        &  14.4 &  24.2 &  20.5   \\
20        &    11.2 &   23.6 &  17.0   \\
30         &    8.9 &    20.3 &   17.2  \\

\bottomrule
\end{tabular} \\
\caption{Valley filter capability $Q_{K^{'}}^u$ when the value of $u=A,b,d$ 
is reduced/increased ($\downarrow/\uparrow)$  by 10, 20 and 30\%. The 
unperturbed (0\%)  parameters corresponds to the superlattice shown in 
\Fref{figure3}(c) $\alpha =14\%$ ($A =  1.17~\text{nm}$, $b=3.12~\text{nm}$, $d 
= 12.79~\text{nm}$ and $N_G = 15$).
}
\label{tab:Q}
\end{table}
 

\begin{figure}[t]
\centering
\scalebox{0.35}{\includegraphics{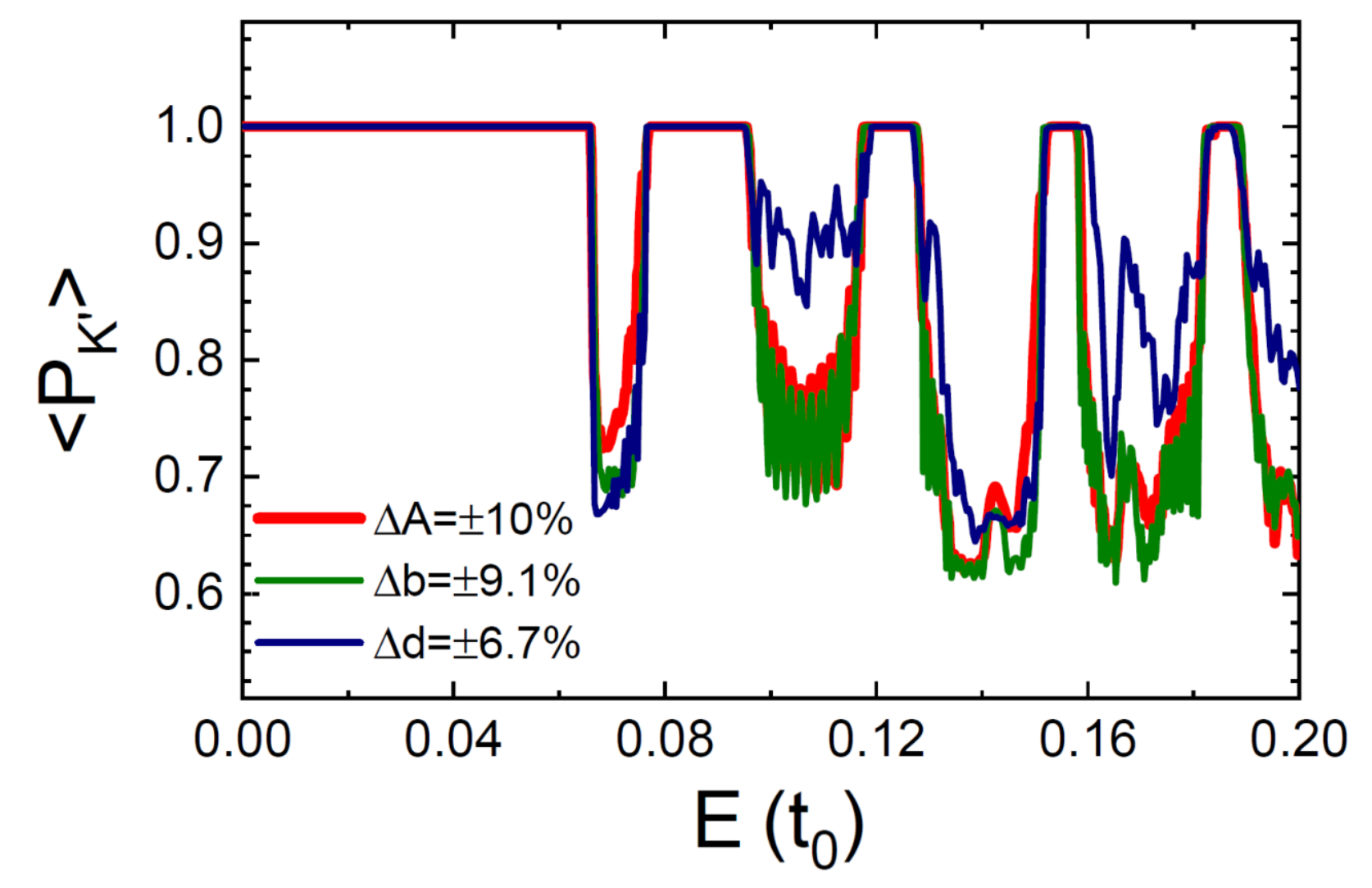}
}
\caption{ Average Valley polarization $<P_{K'}>$ for the 1D Gaussian 
supperlattice presented in \Fref{figure3}(c) $\alpha =14\%$ ($A =  
1.17~\text{nm}$, $b=3.12~\text{nm}$, $d = 12.79~\text{nm}$  and $N_G = 15$) with 
$A$ (height) disorder (red line), $b$ (width) disorder (green line) and $d$ 
(distance) disorder (blue line).  The calculated valley filter capabilities for 
the disordered systems are: $Q_{K^{'}}^A = 21.7\%$, $Q_{K^{'}}^b = 22.7\%$ and 
$Q_{K^{'}}^d = 20.7\%$. 
}
\label{fig:disAbd} 
\end{figure}

 
 Another important point to tackle is the robustness of the valley filter 
against periodicity perturbations. We consider the effect of disorder in 
$A$, $b$ and $d$  by modifying, for each one of the Gaussians in the superlattice, the 
value of $u=A,b,d$ according to $u = u_0 + \delta u$. Again, the unperturbed 
values 
($u_0$) are the parameters of the 1D Gaussian chain with $\alpha = 14\%$ ($A_0 = 
 1.17~\text{nm}$, $b_0=3.12~\text{nm}$,  $d_0 = 12.79~\text{nm}$ and $N_G = 15$) 
and $\delta u$ is a random number uniformly distributed between $[-\Delta u, 
\Delta u]$. Specifically, we fixed $\Delta A = 0.1A_0$, $\Delta b = 0.091b_0$ 
and $\Delta d = 0.067d_0$; the values of $\Delta b$ and $\Delta d$ were chosen 
to avoid overlapping and  deformations none  wider than the width of the 
nanoribbon ($W_0$). The valley polarization averaged   over 20 disorder 
realisations for $A$ (red line), $b$ (green line) and $d$ (blue line) is shown 
in \Fref{fig:disAbd}. It can be seen that the valley filter effect is slightly 
modified by disorder. Besides that, the valley filter capabilities of the 
disordered superlattices 
$Q_{K^{'}}^A = 21.7\%$, $Q_{K^{'}}^b = 22.7\%$ and 
$Q_{K^{'}}^d = 20.7\%$ follow the same trends and have the same 
physical explanation of 
the valley filter capabilities  of the pristine superlattices calculated and 
discussed 
in the previous paragraph.

\section{Modeling Valley Transport using Artificial Neural Networks}

\begin{figure}[t]
\centering
\scalebox{0.75}{\includegraphics{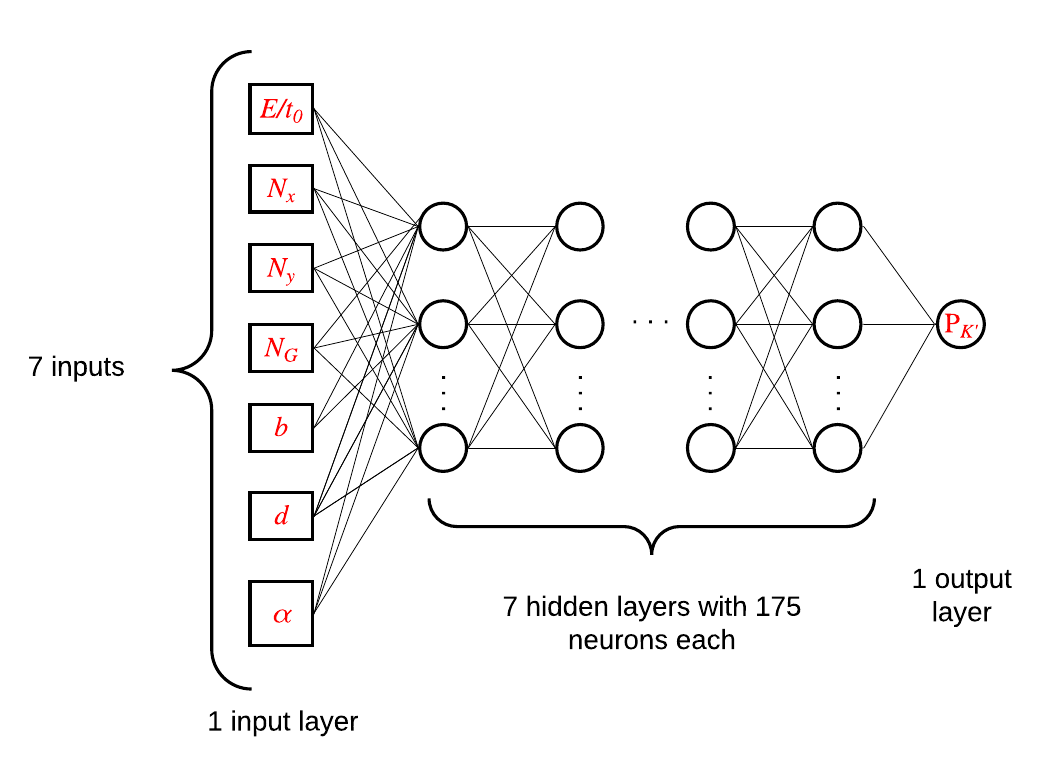}}
\caption{Feedforward fully connected deep neural network architecture used 
for predicting valley polarization. The network consists of: an input layer 
with seven neurons; seven hidden layers with one hundred seventy five neurons 
each and one output layer with a single neuron. The neurons have hyperbolic 
tangent activation with exception of output layer neuron that has linear 
activation.}
\label{figure5} 
\end{figure}

To take advantage of the enhanced valley filter effect produced by the 
Gaussian deformations, it desirable to know in advance the number, energy and 
width of the valley filter plateaus for a given supperlattice. In  this section 
we deal with this problem using Artificial Neural Networks (ANNs). ANNs are 
computational modeling tools that have found extensive acceptance in many 
disciplines for modeling complex real-word problems 
\cite{haykin2010neural,aggarwal2018neural}. The layered architecture of ANN  
(input, hidden and output) is  comprised of densely interconnected adaptive 
simple processing elements called artificial neurons, when a network has more 
than two/three hidden layers it is also called deep neural network (DNN) 
\cite{artificialNeuralNetwork}. The neurons are connected by synaptic weights, 
during the training process the weights are incrementally adjusted and thus the 
network can efficiently learn to perform a specific task. In supervised 
learning, the network is provided with a correct answer (output) for every input 
pattern from the training dataset, Error-Correction Learning (ECL) rule is 
employed to gradually reduce the overall network error. This means that the 
arithmetic difference (error) between the ANN solution at each stage (cycle) 
during training and the corresponding answer is used to modify the synaptic 
weights. The training of ANN occurs from input patterns of the training dataset 
in an interactive process: (i) A sample of this training dataset is randomly 
chosen and introduced in the input layer. (ii) The sample is propagated to the 
output layer in a linear combination of data and weight vectors. (iii) The 
difference between the output of the neural network and the  dataset is used to 
adjust the weight vector. The back-propagation is responsible for propagating 
the error which can be considered as the loss function to be minimized. The 
weights values are then updated according to the gradient descent in a way that 
the total loss is reduced, and a better model is obtained. These steps are 
repeated until the network reaches either a minimum error or a finite number of 
iterations. To assess the performance of a trained model, it is common to use 
the coefficient-of-determination, $R^2$, representing the agreement between the 
predicted and target outputs. Other more involved methods for monitoring network 
training and generalization are based on information theory 
\cite{informationTheory}.

One advantage of using DNNs to model valley polarization is their ability to 
incorporate all the operating parameters in one model; we  assume that  valley 
polarization  may be expressed as unknown function $P_{K'} = f(E/t_0, N_x, N_y, 
N_G, b, d, \alpha)$ of the number of atoms in the $x$ axis direction $N_x$, the 
number of atoms in the $y$ axis direction and the previously defined variables  
$E/t_0$, $N_G$, $b$, $d$, and $\alpha$; in this way, the input  and output 
layers are defined from prior knowledge of the problem (see \Fref{figure5}). It 
is important to highlight that  some of the inputs are highly correlated, but 
even though we decide to include them because they are design parameters of the 
superlattice, on the other hand, highly correlated inputs do not present any 
threat to the training and validation process of the DNN. To determine the ideal 
number of the hidden layers is one of the most critical tasks; several rules are 
available in the literature including those that relate the hidden layer size to 
the number of neurons in input and output layers 
\cite{boger1997knowledge,berry2004data}. However, the high non-linearity of the 
problem forces us to vary the number of hidden layers and their neurons using 
the training error as a decision criterion.  The final DNN architecture has 
seven hidden layers with one hundred seventy-five neurons each as shown in 
\Fref{figure5}.

\subsection{Training and validation}
\begin{table}[b]
\centering
\begin{tabular}{c|c|c|c|c}
\toprule
\textbf{Parameter} &     \textbf{Avg.} &     \textbf{Std.} &   \textbf{Min.} &   
   \textbf{Max.} \\
\midrule
$N_x$        &  1257.76 &  769.01 &  40.00 &  3601.00  \\
$N_y$        &    73.81 &   20.95 &  20.00 &   150.00  \\
$N_G$        &    12.14 &    8.62 &   0.00 &    60.00 \\
$b$            &     2.97 &    0.77 &   0.00 &     4.54  \\
$\alpha$        &     0.16 &    0.12 &   0.00 &     0.40  \\
$d$           &    11.16 &    3.29 &   0.00 &    22.13  \\
\bottomrule
\end{tabular} \\
\caption[Statistical dataset summary of the superlattice 
parameters]{Statistical dataset summary of the superlattice design parameters 
including the  average, standard deviation and minimal/maximal for each 
variable.
}
\label{tab:datasetsummary}
\end{table}

\begin{center}
\begin{figure}[t]
\centering
\scalebox{0.28}{\includegraphics{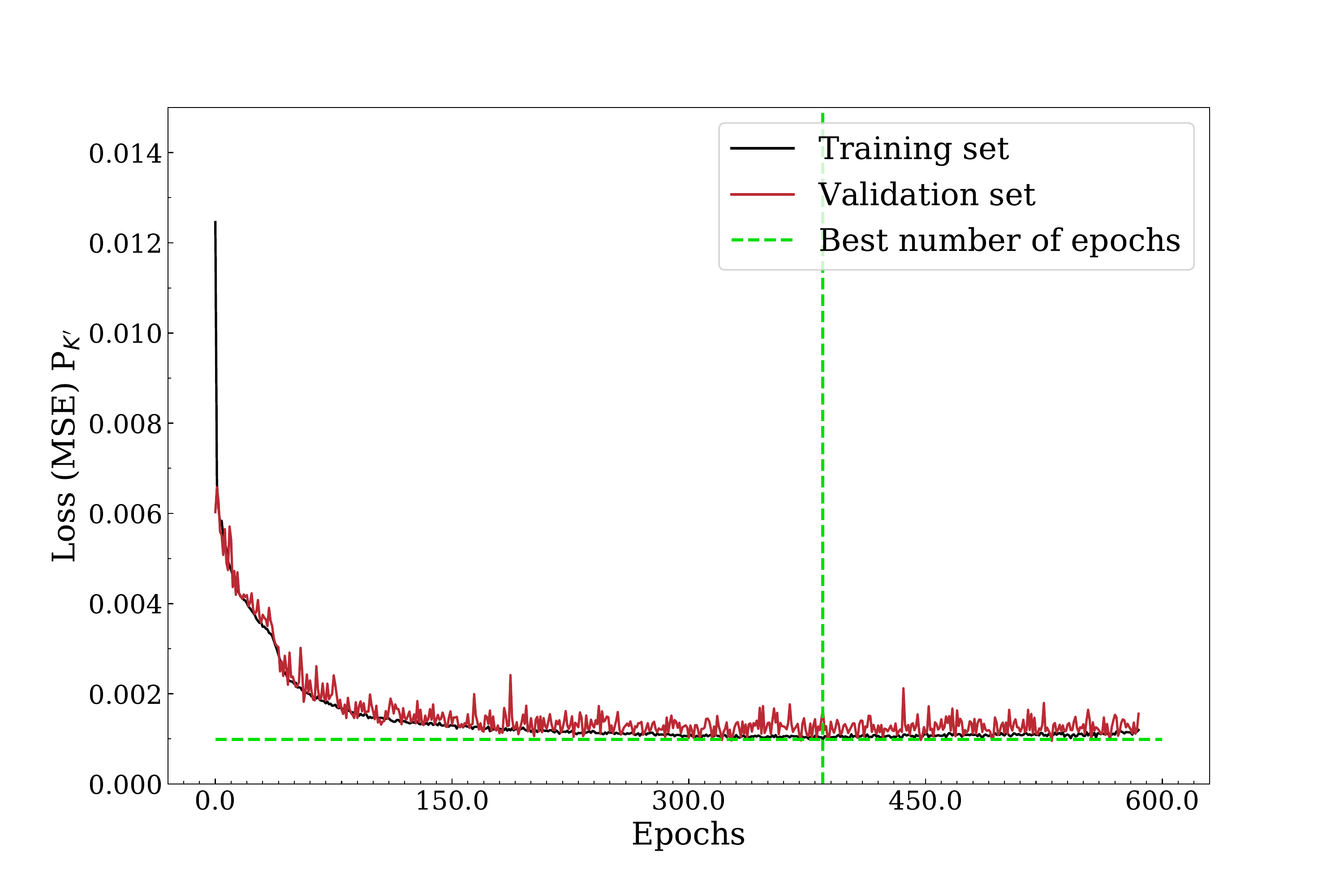}}
\caption{Variation of the MSE versus number of epochs for training and 
test data.} 
\label{figure6} 
\end{figure}
\end{center}

The dataset used in this experiment was generated by the wave function matching 
technique and the Green's Function for 117 distinct configurations of the 1D 
Gaussian supperlattice. Individual configurations defined by the 6-tuple $(N_x, 
N_y, N_G, b, d, \alpha)$ were labeled with the calculated $P_{K'}$ for each one 
of the 500 energy points ($E/t_0 \in [{0.001,0.5}]$). The dataset summary 
presented in table \ref{tab:datasetsummary} helps not only to analyse the 
statistical distribution of the configuration space involved in the ANN 
training, but it also helps understanding how this problem could be modeled, in 
this case, as supervised regression. Training an ANN is typically conducted 
separating the dataset into two sets: training and testing. In this way, one 
trains the model on the training data and then evaluates the performance of the 
model on the testing dataset. It is important that the two sets are 
non-intersecting (i.e. no test input pattern appear in the training dataset) so 
that a fair evaluation of the model generalization is obtained. We randomly 
split the 58500 ($117\times 500$) data samples into 80\% for training and 20\% 
for testing; in addition, 20\% of training dataset was used for hyperparameter 
tunning (i.e changing the number of neurons or layers). Training of the model 
has carried out with a custom TensorFlow \cite{abadi2016tensorflow} 
implementation. The input and output data were preprocessed by z-normalization 
into vectors whose mean is approximately 0 while the standard deviation is in a 
range close to 1 to satisfy the transfer function and to make the training 
faster\cite{haykin2010neural,aggarwal2018neural}.  During training, the loss 
function is monitored and terminated when convergence is obtained. In 
\Fref{figure6} we plot the number of epochs \textit{vs} MSE (Mean Squared Error) 
for training and validation datasets. As the number of epochs increases, the MSE 
is reduced because the weights are updated after each iteration. The loss is too 
high on early epochs, so the model is still underfitting both the training and 
the validation set. Ordinarily, a neural network learns in stages, moving from 
the generalization of simple to more complex mapping functions as the training 
session progresses. This is exemplified by the typical situation in which the 
MSE decreases with an increasing number of epochs during training: it starts off 
at large values, decreases rapidly, and then continues to decrease slowly as the 
network makes its way to a local minimum on the error surface. We identified the 
onset of overfitting through the use of cross-validation, for which the training 
data are split into an estimation subset and a validation subset. The estimation 
subset is used to train the network as usual, but the training session is 
stopped periodically, every so many epoch, and the network is tested on the 
validation subset after each period of training. This simple, effective, and 
widely used approach to training neural networks is called early stopping 
\cite{earlystopping}. The training was stopped at epoch 385, at this stage we 
are sure that the DNN is able to learn from the training set just enough to be 
able to generalize with the validation set ensuring good performance on unseen 
data such as the testing set. Otherwise, overtraining, which refers to exceeding 
the optimal time of ANN training, could result in worse predictive ability of 
the network. As both training and validation loss decrease at a similar rate, 
the best model weights could be found when validation loss stopped improving. 
From epoch 385, the model would stop generalizing and start learning the 
statistical noise in the training dataset. The results using DNN for the 
prediction showed accuracy measures in: $R^2$ (test data) = $0.970$ and MSE 
(test data) = $0.001$. 

 As shown previously, when we do a single evaluation on our test set we get 
only one outcome; this may be the result of some unknown bias. With this in 
mind, we decided to leverage another  important cross-validation technique; 
k-fold assess how the results of statistical analysis and/or the machine 
learning model generalize  independent data sets. To take advantage of this 
technique, the training dataset was partitioned into ten equal-sized subsamples. 
A single subsample was retained as the validation data for testing the model, 
and the remaining nine subsamples were used to train the model. The 
cross-validation was repeated ten times, with each of the ten subsamples used 
exactly once as the validation data.  In order to preserve the target variable 
($P_{K'}$) distribution in training and validation sets, the continuous target 
variable was represented as a categorical variable. To do so, ten buckets/bins 
$[{0.5, 0.55}]$, $[{0.56,0.61}]$,...,$[{0.94, 1}]$ were defined according to the 
value of $P_{K'}$. The ten rounds performance metrics, as shown in the table 
\ref{tab:kfold}, were collected and averaged to produce a single estimation. 
Each round also shows the performance metrics for the trained model predictions 
on the testing dataset. As mentioned previously, when we create ten different 
models and test it on ten different validation sets. By training ten different 
models we can understand better what’s going on. The best scenario is that the 
DNN accuracy is high and the error is low and similar in all ten splits. This 
means that our model and our data are consistent and we can be confident that by 
training it on all the data set and using it in other real-world scenarios will 
lead to similar performance.

\begin{table}[b]
\begin{tabular}{ccccccc}
\toprule\textbf{\#} & \textbf{T. MSE} & \textbf{T. R2} & \textbf{V. MSE} 
& \textbf{V. R2} & \textbf{TS. MSE} & \textbf{TS. R2} \\
\midrule
1            & 0.002           & 0.926          & 0.026           & 0.917        
  & 0.001            & 0.959           \\
2            & 0.002           & 0.925          & 0.031           & 0.910        
  & 0.002            & 0.946           \\
3            & 0.002           & 0.926          & 0.031           & 0.909        
  & 0.002            & 0.947           \\
4            & 0.002           & 0.926          & 0.026           & 0.922        
  & 0.002            & 0.952           \\
5            & 0.002           & 0.925          & 0.025           & 0.921        
  & 0.002            & 0.945           \\
6            & 0.002           & 0.921          & 0.025           & 0.928        
  & 0.002            & 0.950           \\
7            & 0.002           & 0.926          & 0.027           & 0.917        
  & 0.001            & 0.959           \\
8            & 0.002           & 0.925          & 0.027           & 0.915        
  & 0.002            & 0.954           \\
9            & 0.002           & 0.924          & 0.026           & 0.920        
  & 0.002            & 0.951           \\
10           & 0.002           & 0.924          & 0.027           & 0.913        
  & 0.001            & 0.955           \\
\midrule
\textbf{AVG} & 0.002           & 0.925          & 0.027           & 0.917        
  & 0.002            & 0.952           \\
\bottomrule
\end{tabular}
\caption[K-fold performance metrics]{Performance metrics for training 
(T.), validation (V.) and test (TS.) dataset per round of K-fold cross 
validation.}
\label{tab:kfold}
\end{table}

\subsection{Optimal 1D Gaussian Chain}

\begin{figure}[t]
\centering
\scalebox{0.32}{\includegraphics{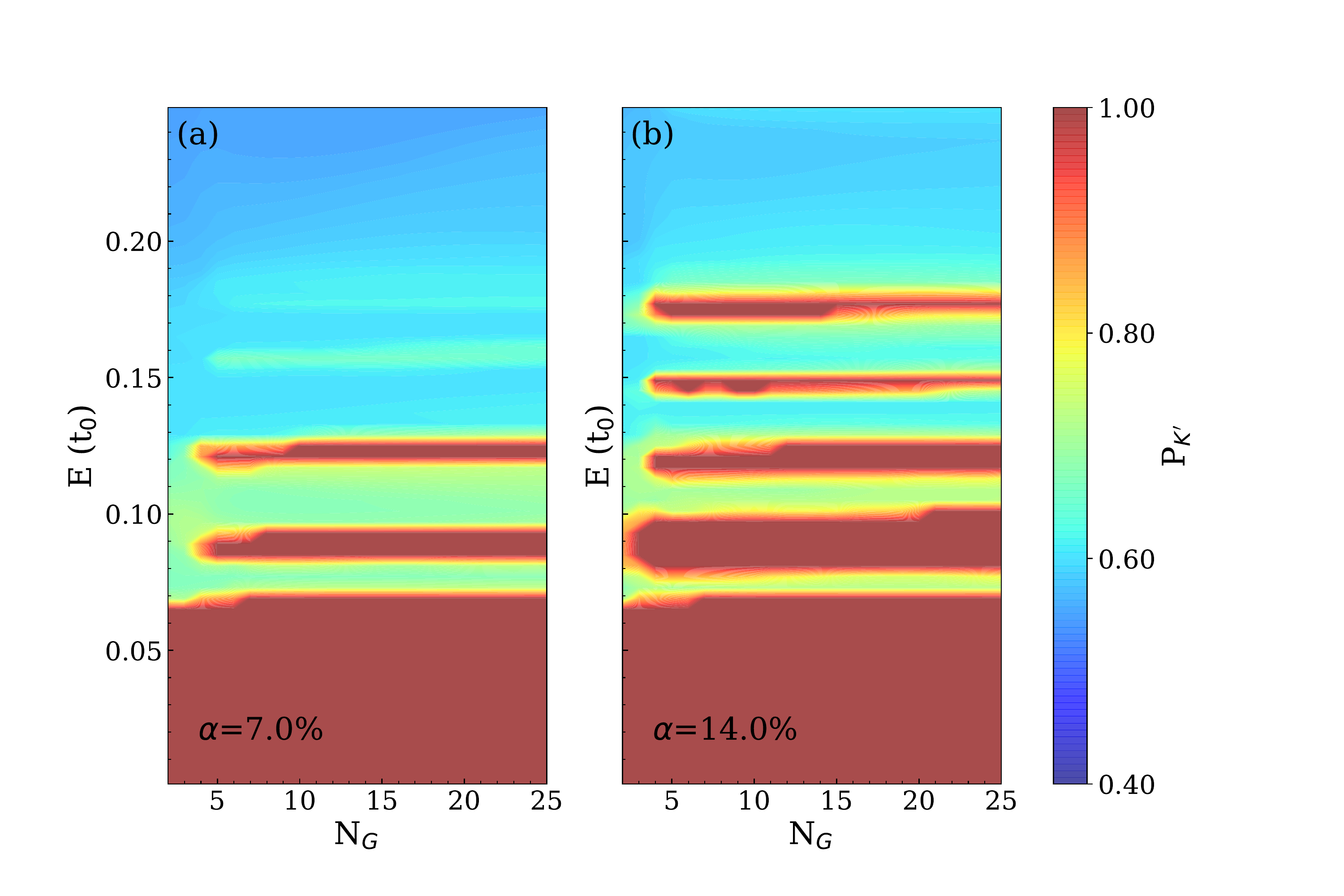}}
\caption{Predicted valley polarization $P_K'$ as function of  E(t$_0$) and 
number of gaussians N$_G$ for (a) $\alpha = 7\%$ and (b) $\alpha = 14\%$. The 
dataset was generated with $N_y = 70~(W_0=14.8~\text{nm})$, $b = 
3.12~\text{nm}$ and $d = 12.7~\text{nm}$} 
\label{figure7}
\end{figure}

In the previous paragraph using data-analysis tools, we show that  the DNN is 
able to learn the relationship between the design parameters of the 1D Gaussian 
chain and  valley polarization. For a physics-related demonstration of the 
performance of the DNN on unseen scenarios, $P_{K'}$ is calculated as function 
of $N_G$ for $N_y = 70~(W_0\approx 14.8~\text{nm})$ , $b = 3.12~\text{nm}$ and $\alpha = 
7\%~\text{and}~14\%$; the  results are plotted in \Fref{figure7}. First, we 
observe the DNN learned that the zigzag edge states plateau is always present, 
irrespective of the values of $N_G$ and $\alpha$. Second, a comparison between 
\Fref{figure7}a and \Fref{figure4} or \Fref{figure3}b for $\alpha = 7\%$, and, a 
comparison between \Fref{figure7}b and  \Fref{figure3}c for  $\alpha = 14\%$ 
clearly show that the DNN predicted  the energy position of the valley plateaus 
as function of $N_G$ and  $\alpha$. In addittion,  \Fref{figure7} makes explicit 
that few Gaussian deformations mimic the effect of an infinite deformation 
chain; in section IIB using the Green’s functions we mentioned that could be the 
case, but now with the DNN this becomes evident with much less computational 
effort.

\begin{figure}[t]
\centering
\scalebox{0.3}{\includegraphics{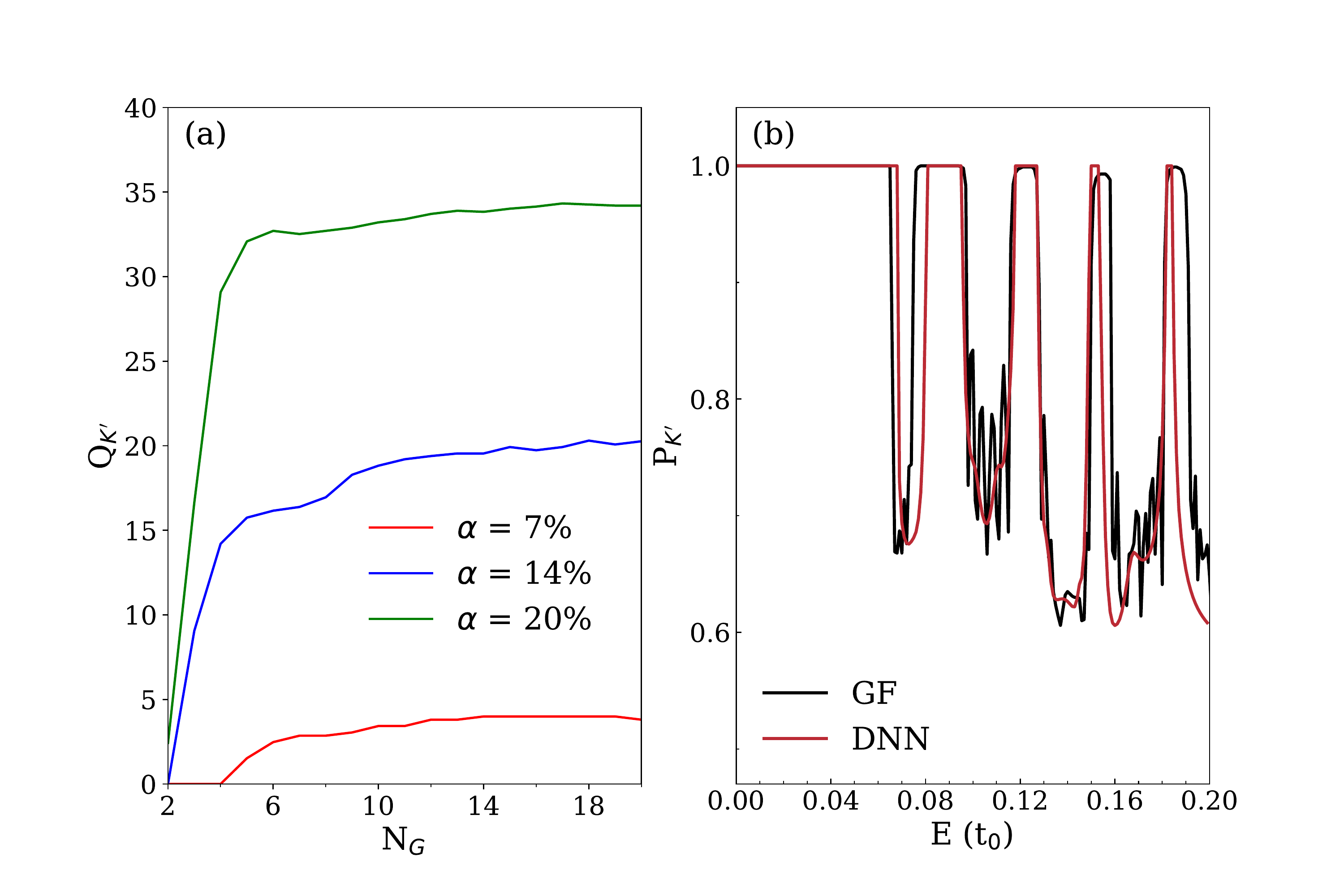}}
\caption{(a) $Q_{K^{'}}$ as a function of $N_G$ for different values of 
$\alpha$ calculated by the DNN. (b) Valley polarization calculated by the 
Green's function (GF) and the DNN for the optimal graphene superlattice 
parameters $N_G = 6$, $\alpha = 15\%$, $N_y = 70~(W_0=14.8~\text{nm})$, $b = 
3.12~\text{nm}$ and $d = 12.7~\text{nm}$.
}
\label{figure9}
\end{figure}

The DNN reproduces the electronic valley transport properties of the 1D 
Gaussian superlattice with an accuracy close to the Green's functions technique. 
This new tool can be used to quickly visualize the relationships among the 
strain superlattice design variables  ($N_x, N_y, N_G, b, d~\text{and}~\alpha$)  
and their effect on valley transport. One example is shown in \Fref{figure9}a 
where the valley filter capability $Q_{K^{'}}$ is calculated as function of 
$N_G$ for $\alpha = 7,14~\text{and}~20\%$. Once more it is seen that  a small 
number of Gaussian deformations in a series configuration enhance valley filter 
capabilities. However, we noticed that $Q_{K^{'}}$ predicted by DNN are smaller 
than the  true values calculated by the Green's functions. The DNN accurately 
predicts the energy location of the valley plate but fails to estimate its 
width. We can also use the trained DNN as a superlattice design tool by 
searching for the 1D Gaussian superlattice with the largest $Q_{K^{'}}$ and the 
smallest strain ($\alpha$) and number of Gaussians ($N_G$). To restrict the 
number of degrees of freedom in the problem but without loss of generality we 
fix the width of the nanoribbon $N_y = 70~(W_0=14.8~\text{nm})$,  $b = 
3.12~\text{nm}$ and  $d = 12.7~\text{nm}$.
The DNN calculated $P_{K'}$ for  696 distinct values of $\alpha \in [{0.01, 
0.3}[$ and $N_G \in [{2,26}[$, then the generated dataset was  sorted by 
$Q_{K^{'}}$ in descending oder. According to our criterion, the optimal 
superlattice corresponds to the system with parameters $N_G = 6$ and $\alpha 
=15\%$. In \Fref{figure9}b we present $P_{K'}$ calculated by the DNN and  the 
Green's function for the optimal case, it is observed that the DNN predicts the 
valley filter for all the energy position and width of the valley plateaus with 
acceptable accuracy. Given that our main objective is the study of the valley 
filter effect we trained the DNN as a model that approximates $P_{K'} \sim 1$ as 
good as possible, by virtue of this choice the DNN does not reproduce the valley 
polarization oscillations for $P_{K'} < 0.8$. Finally, we want to stress that 
the optimal configuration meets our searching criteria, but other criteria can 
be established. For example, one can search for the valley filter with the 
largest number of Gaussians and the smallest strain. The important point is that 
DNN accurately and rapidly calculates $P_{K'}$ and can be used as a design tool 
for highly efficient valley filters.

\section{Outlook and Conclusions}
%
The manipulation of the valley quantum number requires  first effective ways to 
polarise the current, second,  valley low loss transmission systems,  and 
finally valley  detection mechanisms.   The proposed 1D Gaussian strain 
superlattice in a narrow  channel  of  a Field-Effect Transistor  can be used to 
inject or to detect valley polarized currents in a broad range of gate voltages. 
The key ingredient is a superlattice with a period smaller than the electron 
mean free path to induce the folding of the bands and the coupling of 
counter-propagating modes. To achieve this,  graphene can be deposited on 
nanopatterned 
substrates\cite{Jiang:2017hl,Gill:2015fp,Hinnefeld:2018fv,banerjee2019strain, 
Zhang:2018fk}; local measurements of the electronic properties have confirmed 
the appearance of pseudo-Landau levels in the strained regions  providing direct 
evidence of the formation of strain superlattices\cite{Jiang:2017hl}. 
Concretely,   two approaches can be used to produce the studied 1D superlattice. 
(i) Recent works have  transferred graphene to substrates with distinct arrays 
of nanopillars\cite{Jiang:2017hl,C8NR00678D}, the  strain  profile of   graphene 
 on a single  pillar yields a six-fold symmetry pseudo-magnetic field  similar to 
the one generated by one  Gaussian out-of-plane 
deformation\cite{PhysRevB.86.041405}. (ii) Control of the voltage on atomic 
force microscopy (AFM) tips creates  graphene bubbles with a Gaussian profile at 
defined locations\cite{Nemes-Incze,Pengfei20019}.  

On the other hand, our results show that  the valley polarization  of the 
current is robust against 
variations of the strain superlattice parameters ($A$, $b$ and $d$), thus   
graphene does not need to adhere perfectly to all nanopillars, or that all 
graphene Gaussian bubbles obtained by the AFM tip have to have the same height 
and width.  It is important to note that in our transport calculation we do not 
include the effect of wrinkles or inhomogeneous variations of the gate 
potential. However, we do not expect large deviations from our results when 
these effects are taken into account, given that graphene folds are better 
valley filters than a single Gaussian deformation \cite{PhysRevB.98.165437} and 
the small height variations of the bubbles \cite{Zhang:2018fk}.

In summary, we presented a combined Green's functions and Machine learning 
approach to study the valley filter effect in a 1D Gaussian strain  
superlattice. In the first part of this work, using the Green's function and the 
wave function matching technique, we showed that Gaussian superlattice offers 
superior valley filter capabilities compared with the ones observed with a 
single Gaussian out-of-plane deformation. Plotting the band structure of the 
superlattice and identifying the transverse modes, we found that the valley 
filter appears as the result of the coupling between counter-propagating modes 
in the same valley while the strength of the scatterer (value of the 
pseudo-magnetic field) determines the width of the plateau. Although the Green's 
function is solved recursively, the computational complexity of inverting a 
matrix is $\mathcal{O}(N_y^\gamma)$  where $2.3 \leq \gamma \leq 3$ depending on 
the used algorithm, this fact allows us  to explore only a few points in the 
6-dimensional configuration space of the design parameters ($L_0, W_0, N_G, b, 
d~\text{and}~ \alpha$) of the superlattice. With this in mind, in the last 
section we trained a  DNN  to reproduce $P_{K'}\sim 1$; our results show that 
DNN can be used to predict electronic transport properties with an accuracy 
close to the Green’s functions technique but with much less computational effort 
and processing time. 

.
 
\section{Acknowledgments}
VT acknowledges FAPESP under grant 2017/12747-4 and FAPERJ 202.322/2018. PS 
received partial support from Coordena\c{c}\~ao de Aperfei\c{c}oamento de 
Pessoal de N\'ivel Superior - Brasil (CAPES) - Finance Code 001. EATS and DAB 
acknowledge support from FAPESP (process nos. 2012/50259-8, 2015/11779-4 and 
2018/07276-5), the Brazilian Nanocarbon Institute of Science and Technology 
(INCT/Nanocarbono), CNPq and Mackpesquisa.

\bibliography{valley_ML_resub}

\end{document}